\documentclass[RNAAS]{aastex62}

\usepackage{amsmath,amssymb}
\usepackage{bm}

\renewcommand{\vec}{\bm}
\newcommand{\mat}[1]{\bm{\mathrm{#1}}}
\newcommand{\Ev}{\operatorname{E}}
\renewcommand{\Re}{\operatorname{Re}}
\newcommand{\LS}{\mathrm{LS}}
\newcommand{\T}{\mathrm{T}}

\begin{document}

\title{Least Squares Two-Point Function Estimation}

\correspondingauthor{Nicolas Tessore}
\email{nicolas.tessore@manchester.ac.uk}

\author[0000-0002-9696-7931]{Nicolas Tessore}
\affiliation{%
Jodrell Bank Centre for Astrophysics,
University of Manchester,\\
Alan Turing Building,
Oxford Road,
Manchester, M13 9PL, UK}


\section{} 

For a homogeneous and isotropic (i.e.\@ wide-sense stationary) zero-mean random field~$X$ with one set of observed values $x_i$ at distinct locations $i = 1, 2, \ldots$, the estimator for the two-point function (autocovariance function), here denoted~$\xi$, is usually of the form
\begin{equation}\label{eq:1}
	\hat{\xi}(\Delta)
	= \frac{\sum_{d_{ij} \in \Delta} w_i w_j \, x_i x_j}{\sum_{d_{ij} \in \Delta} w_i w_j} \;,
\end{equation}
where $\Delta$ is a distance bin, $d_{ij}$ is the distance between the locations of observations $x_i$ and $x_j$ using a given distance function, and $w_i$ is the weight associated with observation $x_i$.
The expectation of the estimator is
\begin{equation}
	\Ev\bigl[\hat{\xi}(\Delta)\bigr]
	= \frac{\sum_{d_{ij} \in \Delta} w_i w_j \, \xi_{ij}}{\sum_{d_{ij} \in \Delta} w_i w_j} \;,
\end{equation}
where $\xi_{ij} \equiv \xi(d_{ij}) \equiv \Ev[x_i x_j]$ is the two-point function for observed values $x_i$ and $x_j$, and the expectation is taken over realisations of the random field with the observed locations kept fixed.
If the distance bin~$\Delta$ is sufficiently narrow, i.e.\@~$\Delta = [d, d + \epsilon)$ for a given distance~$d$ and small~$\epsilon > 0$, the terms~$\xi_{ij}$ in the sum are all approximately equal to~$\xi(d)$, in which case~$\Ev[\hat{\xi}(\Delta)] \approx \xi(d)$ as desired.
I would like to point out that this ``standard estimator'' is the special case of a larger class of least squares two-point function estimators that interpolate the function values.

The estimator~\eqref{eq:1} can be understood as the weighted average of individual two-point function estimates~$x_i x_j$.
Defining the vector~$\vec{y}$ as the linear enumeration of the distinct pairs~$x_i x_j$,
\begin{equation}\label{eq:3}
	\vec{y}
	= \begin{pmatrix}
		y_1 \\ y_2 \\ y_3 \\ \vdots
	\end{pmatrix}
	= \begin{pmatrix}
		x_1 x_2 \\ x_1 x_3 \\ x_1 x_4 \\ \vdots
	\end{pmatrix} \;,
\end{equation}
the expectation of each element of~$\vec{y}$ on its own is the corresponding two-point function, $\Ev[y_1] = \xi_{12}$, $\Ev[y_2] = \xi_{13}$, etc.
The problem is hence to estimate the mean of random vector~$\vec{y}$ under the constraint that elements with the same associated distance~$d_{ij}$ have the same mean.
This is achieved by fixing a set of values~$\xi_n$, $n = 1, 2, \ldots$, of the true two-point function from which the expectation of the elements of~$\vec{y}$ is interpolated,
\begin{equation}\label{eq:4}
	\Ev[y_m]
	\approx \sum_{n} X_{mn} \, \xi_n \;,
\end{equation}
where $X_{mn}$ is any interpolation scheme that is linear in the function values~$\xi_n$.
Overall, the interpolation~\eqref{eq:4} describes a matrix equation,
\begin{equation}\label{eq:5}
	\Ev[\vec{y}]
	\approx \mat{X} \, \vec{\xi} \;,
\end{equation}
and a weighted least squares estimate~$\vec{\hat{\xi}}_\LS$ of the two-point function values can be obtained as the solution of the normal equations,
\begin{equation}\label{eq:6}
	\mat{X}^\T \mat{W} \mat{X} \, \vec{\hat{\xi}}_\LS
	= \mat{X}^\T \mat{W} \vec{y} \;,
\end{equation}
where~$\mat{W}$ is a weight matrix for random vector~$\vec{y}$.
Since the interpolation matrix~$\mat{X}$ is generally ``tall and skinny'' with many more rows than columns (i.e.\@ many more pairs of observed values than interpolated function values), both the matrix~$\mat{A} = \mat{X}^\T \mat{W} \mat{X}$ and the vector $\vec{b} = \mat{X}^\T \mat{W} \vec{y}$ are small, and solving the normal equations~\eqref{eq:6} is cheap.
Furthermore, for diagonal weight matrices, both $\mat{A}$ and $\vec{b}$ can be computed iteratively from the individual rows of~$\mat{X}$ and~$\vec{y}$, and no large matrix needs to be explicitly constructed.
The least squares estimate is unbiased,
\begin{equation}\label{eq:7}
	\Ev[\vec{\hat{\xi}}_\LS]
	= \bigl(\mat{X}^\T \mat{W} \mat{X}\bigr)^{-1} \, \mat{X}^\T \mat{W} \Ev[\vec{y}]
	\approx \bigl(\mat{X}^\T \mat{W} \mat{X}\bigr)^{-1} \, \mat{X}^\T \mat{W} \mat{X} \, \vec{\xi}
	= \vec{\xi} \;,
\end{equation}
as long as the interpolation~\eqref{eq:5} is a sufficiently good approximation of the true two-point function.
Furthermore, the covariance matrix~$\mat{C}$ of a weighted least squares estimate follows from linearity,
\begin{equation}
	\mat{C}
	= (\mat{X}^\T \mat{W} \mat{X})^{-1} \mat{X}^\T \mat{W} \, \mat{\Sigma} \, \mat{W}^\T \mat{X} (\mat{X}^\T \mat{W}^\T \mat{X})^{-1} \;,
\end{equation}
where $\mat{\Sigma}$ is the covariance matrix of the random vector~$\vec{y}$, which must be computed from the variances and expectations of the observed values~$x_i$.
If weights $\mat{W} = \mat{\Sigma}^{-1}$ can be chosen, the covariance matrix simplifies, $\mat{C} = (\mat{X}^\T \mat{W} \mat{X})^{-1}$.

To obtain the standard estimator~\eqref{eq:1} from the least squares estimate, binning of the two-point function values into given distance bins can be used as one possible interpolation scheme~\eqref{eq:4},
\begin{equation}
	X_{mn}
	\equiv \begin{cases}
		1 & \text{if the distance for $y_m$ is in distance bin $n$,} \\
		0 & \text{otherwise.}
	\end{cases}
\end{equation}
In particular, each value then belongs to at most one bin, so that $X_{mn} X_{mn'} \equiv X_{mn} \delta_{nn'}$, where $\delta$ is the Kronecker delta.
For a diagonal weight matrix~$\mat{W}$, the normal equations~\eqref{eq:6} simplify:
The matrix on the left-hand side is diagonal,
\begin{equation}
	(\mat{X}^\T \mat{W} \mat{X})_{nn'}
	= \sum_{m} W_{mm} X_{mn} X_{mn'}
	= \delta_{nn'} \sum_{m} W_{mm} X_{mn} \;,
\end{equation}
and can be brought to the right-hand side by simple division.
The result is precisely the standard estimator~\eqref{eq:1} for the two-point function,
\begin{equation}
	(\vec{\hat{\xi}}_\LS)_n
	= \frac{\sum_{m} W_{mm} X_{mn} y_m}{\sum_{m} W_{mm} X_{mn}} \;,
\end{equation}
which is thus indeed a special case of the least squares estimator using binning instead of interpolation.
However, a true interpolation scheme estimates the function values at specific points, instead of binned averages.

\begin{figure}[htb!]%
\centering%
\includegraphics{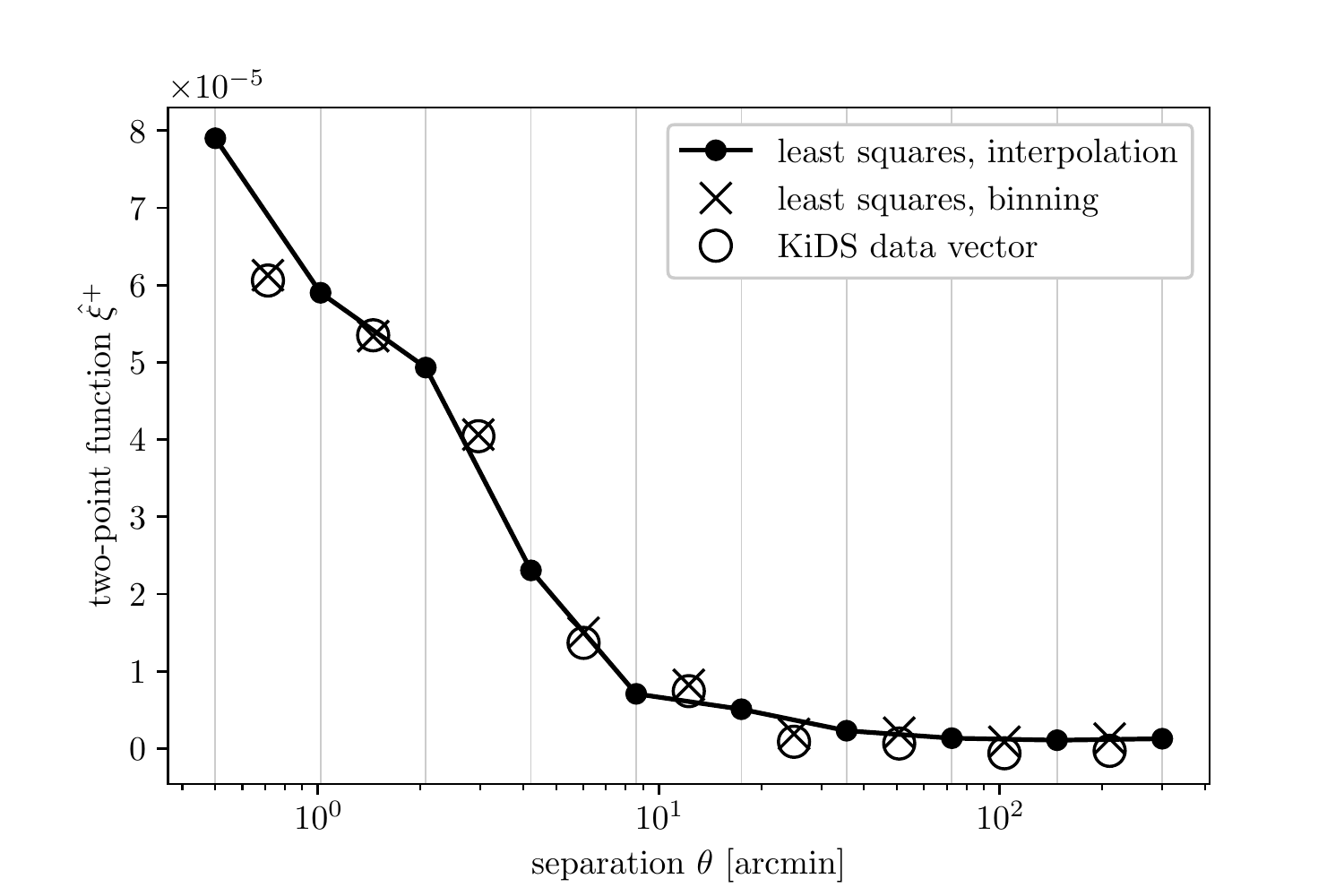}%
\caption{%
Estimated two-point function~$\hat{\xi}^{+}$ for the publicly available \emph{KiDS-450} weak lensing data in the first tomographic redshift bin \citep[cf.\@][]{2017MNRAS.465.1454H}.
Shown are the least squares estimates using log-linear interpolation (\emph{solid line}) and logarithmic binning (\emph{crosses}), as well as the published \emph{KiDS-450} data vector (\emph{open circles}).
The vertical lines indicate the function bins or interpolation points.
}%
\label{fig:1}%
\end{figure}

To demonstrate the least squares estimator, it is applied to a problem in Cosmology, where the two-point function of the shear field~$g$ due to weak gravitational lensing is routinely measured \citep[for a review, see e.g.\@][]{2015RPPh...78h6901K}.
Because of the spin-2 nature of the shear field, the definition of the two-point function~$\xi^{+}$ is slightly more elaborate,
\begin{equation}
	\xi^{+}_{ij}
	= \Re \Ev\bigl[g_i g_j^* \, \mathrm{e}^{-2 \, \mathrm{i} \, \phi_{ij}}\bigr] \;,
\end{equation}
where $\phi_{ij}$ accounts for the relative orientation between $g_i$ and $g_j$.
Figure~\ref{fig:1} shows the least squares estimates for the publicly available \emph{KiDS-450} weak lensing data \citep{2017MNRAS.465.1454H} using both logarithmic bins and log-linear interpolation, as well as the published \emph{KiDS-450} data vector obtained using the \emph{athena} code \citep{2014ascl.soft02026K}.
As expected, the binned least squares estimate corresponds closely to the standard estimator, with small differences likely due to the approximations used in the tree code.
The interpolated least squares estimate is similarly in good agreement.

\acknowledgments

I would like to thank J.~P.~Cordero for help with processing the data.
The author acknowledges support from the European Research Council in the form of a Consolidator Grant with number 681431.

\bibliography{lstpfe}

\end{document}